\documentclass[11pt]{article}

\usepackage[utf8]{inputenc}
\usepackage[T1]{fontenc}
\usepackage[margin=1in]{geometry}
\usepackage{cite}
\usepackage{amsmath,amsfonts,amssymb}
\usepackage{siunitx}
\usepackage{graphicx}
\usepackage{subcaption}
\usepackage{textcomp}
\usepackage{booktabs,multirow,array}
\usepackage[section]{placeins}
\usepackage[table]{xcolor}
\usepackage[hidelinks]{hyperref}
\graphicspath{{fig/}}
\title{Differential Stochastic Simulated Annealing Processor for Fully Connected 2048-Spin Optimization}

\author{%
 Naoya~Onizawa$^{1}$,
 Md~Mohaimenul~Alam$^{1,2}$,
 \\
 Sean~Smithson$^{1,2}$,
 Duckgyu~Shin$^{1,2}$,
 and Takahiro~Hanyu$^{1}$\\[0.8em]
 \normalsize $^{1}$Research Institute of Electrical Communication, Tohoku University, Sendai, Miyagi, Japan\\
 \normalsize $^{2}$Graduate School of Engineering, Tohoku University, Sendai, Miyagi, Japan
}
\date{}

\begin{document}

\maketitle

\begin{abstract}
A 2,048-spin fully connected annealing processor based on differential stochastic simulated annealing (DSSA) is presented as an architectural design in TSMC \SI{28}{\nano\meter} CMOS with a \SI{3}{\milli\meter}\texttimes\SI{4}{\milli\meter} post-layout area. The processor closes timing at \SI{500}{\mega\hertz}, integrates a 16 Mb SRAM weight memory, and amortizes stochastic noise across 16 spins with area-efficient random number generators. DSSA keeps a serialized datapath for density but recomputes interactions only for spins that flip, shrinking the effective workload to the active frontier during each annealing run. Spin-select scheduling, priority-based weight reads, and a temperature controller that skips idle steps accelerate sparse updates without sacrificing full connectivity. Post-layout simulation results show \SI{2.7}{\milli\second} time-to-solution (TTS) and \SI{0.86}{\milli\joule} energy-to-solution on 2,000-spin problems at \SI{316}{\milli\watt} (\SI{0.15}{\milli\watt}/spin), achieving 1.5$\times$ lower power and $3.5\times$ lower TTS energy than projected prior fully connected annealers. These results demonstrate the potential of the proposed DSSA architecture for large-scale combinatorial optimization hardware under post-layout evaluation.
\end{abstract}


\section{Introduction}
Combinatorial optimization problems (COPs) such as MAX-CUT, floorplanning, and routing map naturally to Ising models whose minima encode feasible solutions~\cite{Lucas2014FrontPhys}. These problems are NP-hard in general~\cite{nphard}, and simulated annealing (SA) has long been a standard heuristic~\cite{Kirkpatrick1983Science}.

Fully connected CMOS annealers must either provision $N^2$ weight bandwidth each step or revisit all $N$ spins serially; both routes inflate cycles, area, and power when $N$ scales to the thousands. Wide convolution engines and 
multi-chip tiling handle dense graphs but increase interconnect complexity and energy consumption~\cite{Yamaoka2015JSSC,Aramon2019NC}, while sparse topologies or embeddings lose fidelity for dense problems~\cite{Johnson2011Nature,Albash2018RPP,Takemoto2021ISSCC,Yamamoto2021JSSC,Kawamura2023ISSCC,Chu2024ISSCC}.

Recent fully connected digital annealers illustrate the tension: 512--576 spins at sub-\SI{400}{\mega\hertz} consume hundreds of milliwatts even with multi-chip partitioning~\cite{Yamamoto2021JSSC,Kawamura2023ISSCC}. 
In this paper, one \emph{shot} denotes a complete annealing run over one temperature schedule.{}
For fully connected designs, $N^2$ serialization pushes per-shot latency to thousands of cycles.
Parallel datapaths avoid the latency, but require $N$ multiplexers and $N^2$ weight fetches per step, which becomes prohibitive beyond a few hundred spins. Serial datapaths shrink mux count to $1/N$ of parallel, but 
still require $N$ cycles per step. Therefore, a dense 2,000-spin graph requires
on the order of $2{,}000$ cycles per temperature regardless of how many spins actually change.
Approaches that rely on sparse graphs or embeddings reduce bandwidth, but distort dense problem structure.

This work focuses on the architectural design and post-layout evaluation of a single-chip, fully connected 2{,}048-spin processor that avoids the quadratic cost of naive serialization. Differential stochastic simulated annealing (DSSA) keeps a serialized datapath for density but only recomputes interactions for spins that flip, so the workload scales with the active frontier rather than all spins during a shot. Spin-select scheduling, priority weight reads, and stepwise temperature control inside each shot skip idle steps, while 16:1 RNG sharing and an on-chip 16\,Mb SRAM sustain full connectivity at \SI{500}{\mega\hertz}. Post-layout simulation shows \SI{2.7}{\milli\second} time-to-solution (TTS) on 2,000-spin MAX-CUT with \SI{0.86}{\milli\joule} energy-to-solution at \SI{316}{\milli\watt} (\SI{0.15}{\milli\watt}/spin), 1.5$\times$ lower power and $3.5\times$ lower TTS energy than a projected 2,000-spin STATICA~\cite{Yamamoto2021JSSC}. Contemporary GPU SA baselines consume \SI{52.8}{\watt} at hundreds of milliseconds per anneal on RTX-class devices, underscoring a three-order power gap. Because prior ASIC-based annealers are typically reported with fabricated-silicon validation whereas the proposed DSSA processor is evaluated using post-layout simulation, the comparative results in this paper should be interpreted with that difference in validation methodology in mind.

Unlike conventional time-division implementations that simply serialize an otherwise full spin-update computation, the proposed DSSA processor is based on an algorithm--architecture co-design around a differential-update principle: between consecutive annealing steps, changes in the interaction field arise only from spins that flip, and therefore only the interaction terms affected by those flipped spins are recomputed. The performance gain therefore originates from reducing redundant interaction evaluations based on spin-transition sparsity, rather than from multiplexing alone.

The main contributions are:
\begin{itemize}
 \item A 2,048-spin fully connected DSSA processor architecture in TSMC \SI{28}{\nano\meter} CMOS that closes timing at \SI{500}{\mega\hertz} in post-layout evaluation and achieves \SI{2.7}{\milli\second} TTS and \SI{0.86}{\milli\joule} energy-to-solution on 2,000-spin MAX-CUT at \SI{316}{\milli\watt} (\SI{0.15}{\milli\watt}/spin).
 \item A DSSA algorithm and serializer that touch only flipped spins, cutting serialized stochastic simulated annealing (SSA) cycles by up to $30.6\times$ versus conventional full-spin sweeps while preserving full connectivity and solution quality.
 \item Architectural co-design—spin-select queuing, priority WM banking, 16:1 RNG sharing, and shot-level temperature control—that aligns bandwidth and stochastic resources with sparse activity to reduce area and power.
\end{itemize}

The remainder of the paper is organized as follows. Section~\ref{sec:sa} reviews Ising formulations, SA basics, and related annealers. Section~\ref{sec:dssa} details DSSA and contrasts it with SSA. Section~\ref{sec:architecture} presents the processor architecture and circuit implementation. Section~\ref{sec:evaluation} reports software validation, hardware performance estimates, and benchmark comparisons. Section~\ref{sec:conclusion} concludes.

\section{Preliminaries and Related Work}
\label{sec:sa}
\subsection{Ising Formulation and SA Basics}
Given a search space $F$ and cost function $c: F \rightarrow \mathbb{R}$, COPs seek $f \in F$ satisfying $c(f) \leq c(y)$ for all $y \in F$. MAX-CUT is a canonical example: given a weighted graph, the goal is to partition vertices into two sets that maximize the sum of edge weights crossing the cut. Mapping MAX-CUT to an Ising Hamiltonian yields~\eqref{eq:ising_energy}, where spins $\sigma_i\in\{\pm1\}$ encode the partition:
\begin{equation}
E(\boldsymbol{\sigma}) = -\sum_{i} h_i \sigma_i - \frac{1}{2}\sum_{i, j} J_{ij} \sigma_i \sigma_j,
\label{eq:ising_energy}
\end{equation}

\begin{figure}[htbp]
 \centering
 \includegraphics[width=0.36\columnwidth]{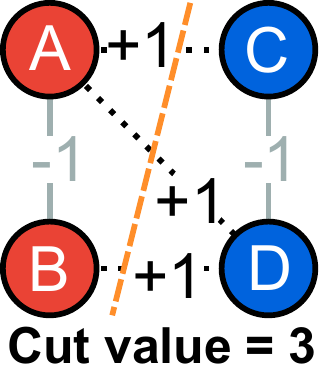}
 \caption{Four-node MAX-CUT illustration highlighting the partition objective.}
 \label{fig:fpga_bench}
\end{figure}

\begin{figure}[htbp]
	\centering
	\includegraphics[width=\columnwidth]{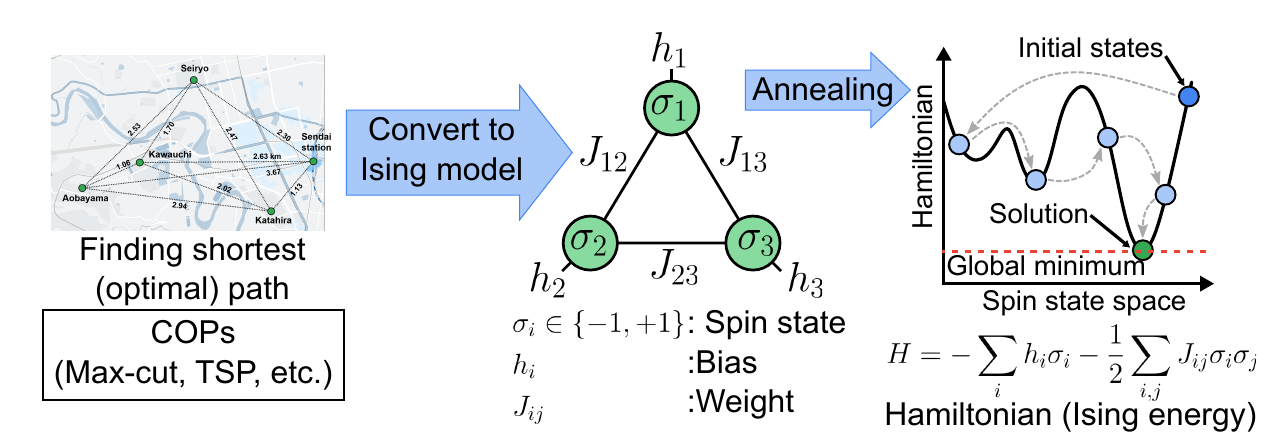}
	\caption{Conversion flow from a MAX-CUT instance to an Ising formulation and its Hamiltonian representation.}
	\label{fig:SA}
\end{figure}

Fig.~\ref{fig:fpga_bench} shows a four-node MAX-CUT toy used here to illustrate the partition objective: the optimum separates nodes into two groups that maximize the cut weight.
As shown in Fig.~\ref{fig:SA}, a weighted graph is mapped to spin variables $\sigma_i$, where edge weights translate to coupling terms $J_{ij}$ and node biases correspond to $h_i$. The resulting Hamiltonian encodes the MAX-CUT objective, enabling evaluation by SA.

SA perturbs the current spin state to $\boldsymbol{\sigma}'$ and accepts with probability
\begin{equation}
 p(\boldsymbol{\sigma}') = \exp\!\left(-\frac{E(\boldsymbol{\sigma}')-E(\boldsymbol{\sigma})}{T}\right),
 \label{eq:sa_accept}
\end{equation}
where $T$ is the temperature. High temperatures accept uphill moves, enabling exploration; cooling schedules such as $T(t{+}1)=T(t)\alpha$ gradually reduce this probability and guide convergence toward the maximum cut.

\begin{figure}[htbp]
	\centering
	\includegraphics[width=\columnwidth]{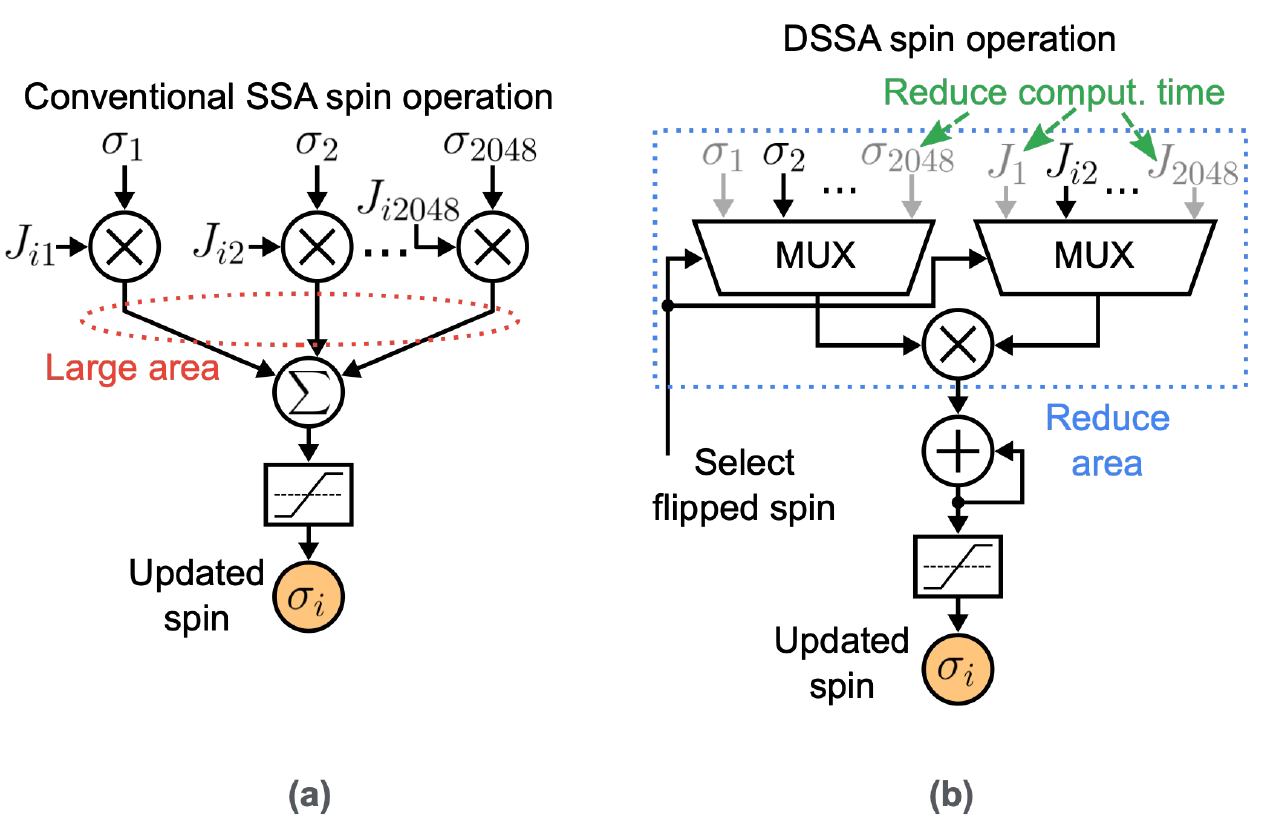}
	\caption{Conceptual comparison between (a) conventional stochastic simulated annealing (SSA) and (b) the proposed differential SSA (DSSA). SSA evaluates all spins at every step, whereas DSSA updates only the spins that flip between consecutive states, reducing redundant computations. }
	\label{fig:ssa_vs_dssa}
\end{figure}

\subsection{Hardware Realizations of SA and QA}
Hardware annealers pursue speedups through different physical substrates. Quantum annealers promise tunneling-based escapes from local minima but require cryogenic qubits and coherent control~\cite{QA,dwave}. Probabilistic bits realized with MTJs offer room-temperature stochastic behavior~\cite{pbit_model,sa_pbit} yet face scaling and integration overhead.

SSA~\cite{SSA} approximates the $\tanh(\cdot)$ nonlinearity with stochastic computing~\cite{sto_book} and has been prototyped on FPGAs~\cite{jetcas}, but sparse topologies (for example, King's graphs) limit fidelity for dense problems.

More recently, sparse parallel-update Ising machines have continued to be actively investigated. For example, an all-to-all reconfigurable sparse and higher-order Ising machine has been proposed to emulate dense interactions while preserving highly parallel spin updates~\cite{Nikhar2024AllToAll}. We therefore regard such sparse-topology machines as an important related design direction, while distinguishing them from the fully connected multilevel-weight target of this work.

ASIC designs extend spin counts~\cite{Takemoto2021ISSCC,Yamamoto2021JSSC,Kawamura2023ISSCC} while trading between connectivity and cost: some rely on multi-chip tiling or wide convolution engines~\cite{Yamaoka2015JSSC,Aramon2019NC}, whereas others serialize updates and revisit all spins each step. These tradeoffs motivate a differential serialized approach that retains full connectivity without quadratic datapaths.

\section{Differential Stochastic Simulated Annealing}
\label{sec:dssa}
\subsection{Area--Latency Motivation}
Fig.~\ref{fig:ssa_vs_dssa} highlights the difference in search trajectories and activity patterns, emphasizing how DSSA shrinks the effective neighborhood while preserving full connectivity.
SSA’s spin interaction can be parallelized or fully serialized. Parallel spin-gates require $N$ multiplexers and update all spins in one cycle but incur substantial area overhead for large $N$; serial spin-gates reuse one multiplexer and cut area to $1/N$ of parallel but require $N$ cycles per annealing step.

%
DSSA retains the serialized datapath but reduces cycles by updating only spins that actually flip. 

The serializer and multiplexer structure should therefore be viewed as an implementation mechanism for exploiting the differential-update principle. In contrast to ordinary time-division serialization, which only reuses hardware while preserving the same amount of per-step computation, DSSA reduces the number of interaction evaluations themselves by scheduling only the flipped-spin contributions.

Table~\ref{tab:t4_area_clk_comp} compares multiplexer count and cycles per step when $N$ spins are present and $M$ of them flip.
\begin{table}[htbp]
 \caption{Multiplexer count and cycles per step for parallel/serial SSA and DSSA with $N$ spins and $M$ flips.}
 \label{tab:t4_area_clk_comp}
 \centering
 \begin{tabular}{l|ccc}
 \hline\hline
 & Parallel SSA & Serial SSA & DSSA \\
 \hline
 Multiplexers per \\ spin-gate & $N$ & 1 & 1 \\
 Cycles per step & 1 & $N$ & $M\;(0\!\leq\! M\!\leq\! N)$ \\
 \hline\hline
 \end{tabular}
\end{table}
Because high inverse temperature suppresses spin flips late in a shot, $M$ is typically far smaller than $N$, allowing DSSA to retain the serial area profile while reducing per-step cycles.

\subsection{Initialization at $t{=}0$}
With no prior state at the first annealing step, every spin executes a full SSA-style update. The input current $I_i(1)$, clipped interaction $\mathrm{Itanh}_i(1)$, and resulting spin $\sigma_i(1)$ follow
\begin{subequations}
 \begin{align}
 I_i(1) &= h_i + \sum_j J_{ij} \sigma_j(0) + n_{\text{rnd}}\, r_i(0), \label{eq:dssa_init_I}\\
 \mathrm{Itanh}_i(1) &=
 \begin{cases}
 I_0(0) - 1, & I_i(1) \geq I_0(0),\\
 -I_0(0), & I_i(1) < -I_0(0),\\
 I_i(1), & \text{otherwise},
 \end{cases} \label{eq:dssa_init_tanh}\\
 \sigma_i(1) &= \mathrm{sgn}\!\left(\mathrm{Itanh}_i(1)\right), \label{eq:dssa_init_sigma}
 \end{align}
 \label{eq:dssa_init}
\end{subequations}
where $\sigma_i(0)$ is the spin initializer (set to $-1$ in hardware to avoid costly searches) and $I_0$ is the pseudo-inverse temperature. 
Here, $r_i(t)\in\{+1,-1\}$ denotes an independent random variable generated at each annealing step, and $n_{\text{rnd}}$ controls the amplitude of the stochastic perturbation injected into the local field, analogous to thermal noise in conventional simulated annealing.{}{} 
This establishes the FSM state for the integral stochastic $\tanh(\cdot)$ used thereafter.

\subsection{Differential Updates for $t{>}0$}
For later steps, only spins that flipped in the previous step raise a differential flag $\Delta_j(t)\!\in\!\{0,1\}$. From the FSM viewpoint, the increment is the difference of consecutive interaction states,
\begin{equation}
 \Delta I_i(t{+}1) = \mathrm{Itanh}_i(t{+}1) - \mathrm{Itanh}_i(t),
 \label{eq:dssa_fsm_move}
\end{equation}
and since $\mathrm{Itanh}_i$ tracks $I_i$ in the stochastic FSM, we approximate
\begin{align}
 \Delta I_i(t{+}1) &= I_i(t{+}1) - I_i(t) \nonumber\\
 &= \sum_j J_{ij}\big[\sigma_j(t)-\sigma_j(t{-}1)\big] + n_{\text{rnd}}\,r_i(t).
 \label{eq:dssa_delta_expand}
\end{align}
The flip flag
\begin{equation}
 \Delta_j(t) =
 \begin{cases}
 1, & \sigma_j(t)\neq \sigma_j(t{-}1),\\
 0, & \text{otherwise},
 \end{cases}
 \label{eq:dssa_flip_flag}
\end{equation}
zeros out unchanged spins. For a flipped spin, $\sigma_j(t)-\sigma_j(t{-}1)$ equals $+2$ if $\sigma_j(t){=}1$ and $-2$ if $\sigma_j(t){=}-1$, so
\begin{equation}
 \sigma_j(t)-\sigma_j(t{-}1) = 2\,\Delta_j(t)\,\sigma_j(t).
 \label{eq:dssa_sigma_diff}
\end{equation}
Substituting \eqref{eq:dssa_flip_flag} and \eqref{eq:dssa_sigma_diff} into \eqref{eq:dssa_delta_expand} yields
\begin{equation}
 \Delta I_i(t{+}1) = 2 \sum_j J_{ij}\,\Delta_j(t)\,\sigma_j(t) + n_{\text{rnd}}\,r_i(t).
 \label{eq:dssa_delta_flag}
\end{equation}
Using \eqref{eq:dssa_delta_flag} to update the stored interaction current gives
\begin{equation}
 I_i(t{+}1) = I_i(t) + 2 \sum_j J_{ij}\,\Delta_j(t)\,\sigma_j(t) + n_{\text{rnd}}\,r_i(t),
 \label{eq:dssa_prov_I}
\end{equation}
so the SSA acceptance rule can be evaluated without recomputing contributions from spins that did not flip. Operationally, DSSA updates for $t{>}0$ follow

\begin{subequations}\label{eq:dssa_ops}
 \begin{align}
 \Delta I_i(t{+}1) &= 2 \sum_j J_{ij}\,\Delta_j(t)\,\sigma_j(t) + n_{\text{rnd}}\,r_i(t), \label{eq:delta_current}\\
 X_i(t{+}1) &\triangleq \mathrm{Itanh}_i(t) + \Delta I_i(t{+}1), \notag\\
 \mathrm{Itanh}_i(t{+}1) &=
 \begin{cases}
 I_0(t) - 1, & X_i(t{+}1) \geq I_0(t), \\
 -I_0(t), & X_i(t{+}1) < -I_0(t), \\
 X_i(t{+}1), & \text{otherwise},
 \end{cases}
 \label{eq:dssa_itanh_clip}\\
 \sigma_i(t{+}1) &=
 \begin{cases}
 1, & \mathrm{Itanh}_i(t{+}1) \geq 0,\\
 -1, & \text{otherwise},
 \end{cases}
 \label{eq:dssa_sigma_update}
 \end{align}
\end{subequations}

which mirror SSA’s clipped $\tanh$ update while reducing work to the $M$ flipped spins ($M\!\leq\!N$) instead of all $N$.

\begin{figure}[htbp]
	\centering
	\includegraphics[width=1.0\textwidth]{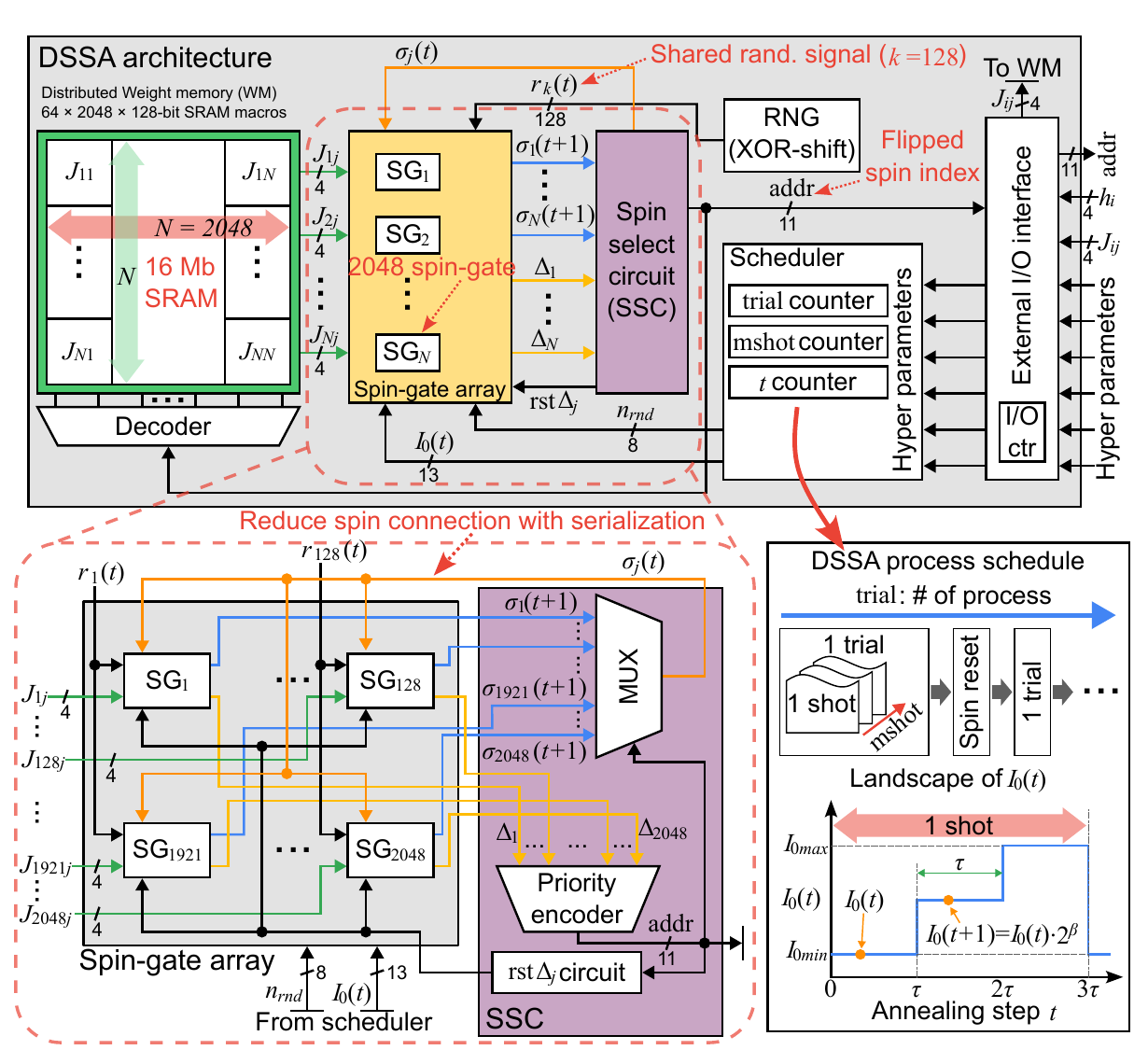}
	\caption{Top-level floorplan and data routing of the DSSA processor. The figure highlights how the RNG banks feed 128 shared bits into the spin-gate array (16:1 sharing across spins), how the weight memory (WM) banks interact with the spin-select circuit (SSC), and how the scheduler broadcasts $I_0$ parameters to enforce the temperature profile in \eqref{eq:temperature}.}
	\label{fig:architecture}
\end{figure}

\subsection{Temperature Scheduling and Acceptance}
The flip decision still follows SSA. The stochastic acceptance uses
\begin{equation}
P\!\left(\sigma_i(t{+}1)=1\right)=\tfrac{1}{2}\left[1+\tanh\!\left(\frac{I_i(t{+}1)}{I_0(t)}\right)\right],
\label{eq:probability}
\end{equation}
where $I_0$ is updated every step as
\begin{equation}
 I_0(s) = I_{0\min} + \left(I_{0\max}-I_{0\min}\right)\left(1-e^{-s/\tau}\right)^{\beta},
\label{eq:temperature}
\end{equation}
with step index $s$ within a shot; one shot spans the ramp from $I_{0\min}$ to $I_{0\max}$. High $I_0$ suppresses flips and keeps $M$ small late in a shot, while early low $I_0$ enables exploration. Hardware implementations of the stochastic nonlinearity, RNG sharing, and workload-aware scheduling are described in Section~\ref{sec:architecture}.

\section{Processor Architecture and Circuits}
\label{sec:architecture}
\subsection{Top-Level Architecture}
The top-level architecture in Fig.~\ref{fig:architecture} translates the DSSA algorithm into hardware. The spin-gate (SG) array integrates 2,048 SGs, each representing a binary Ising state. Random noise signals drive the SGs, with every 16 spins sharing a single 1-bit random signal to minimize random number generator (RNG) area. Consequently, only 128 random bits are generated per cycle, and the shared stream is produced by an on-chip xorshift engine that updates a 128-bit vector every clock. Weights are stored in a 4-bit precision SRAM WM whose 16\,Mb capacity is partitioned into banks aligned with the serializer bandwidth.
Fig.~\ref{fig:architecture} also shows how the spin-select circuit gathers $\Delta_i$ flags from the SG array, prioritizes active spins, and issues WM addresses so that only flipped spins trigger 4-bit weight reads, while the scheduler broadcasts the shot-level $I_0$ parameters to every SUC in lockstep.
Although Fig.~\ref{fig:architecture} illustrates the weight memory as a single logical block for clarity, the 16\,Mb weight memory is physically implemented using 64 distributed SRAM macros, each with a capacity of 2048 words $\times$ 128 bits. These macros are placed around the spin-gate array and are accessed in parallel through the serializer and scheduler. From the viewpoint of the spin-gate array, the distributed SRAM macros collectively function as a single logical weight memory.{}

\begin{figure}[htbp]
	\centering
	\includegraphics[width=1.0\textwidth]{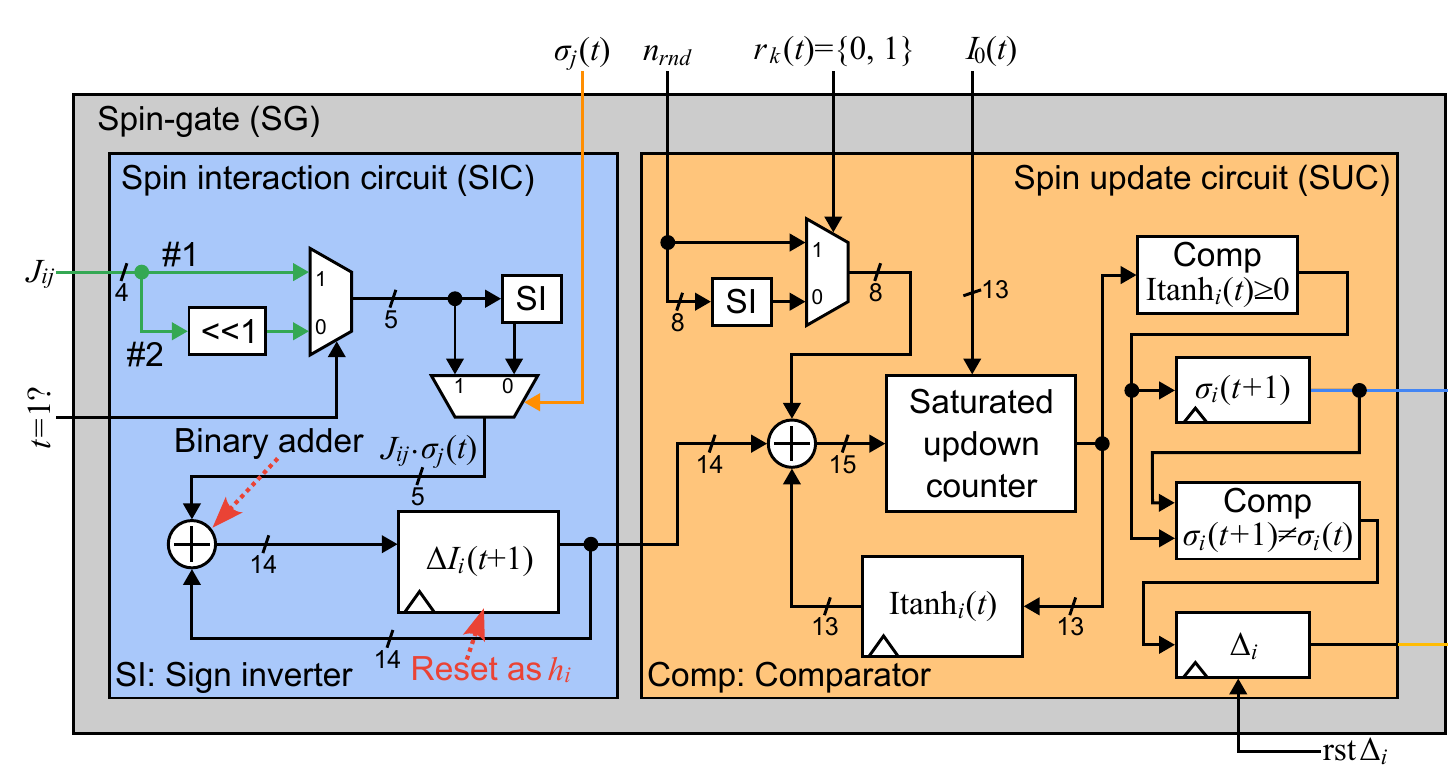}
	\caption{Detailed spin-gate (SG) circuit indicating SIC accumulators for \eqref{eq:delta_current}, SUC finite-state machines that realize the probability in \eqref{eq:probability}, and the annealing timeline showing initialization, selective updates, and SSC draining. The inset emphasizes that only SGs with $\Delta_i=1$ request WM entries, which keeps current draw proportional to the number of flips.}
	\label{fig:spingate}
\end{figure}

\begin{figure}[htbp]
	\centering
	\includegraphics[width=1.0\textwidth]{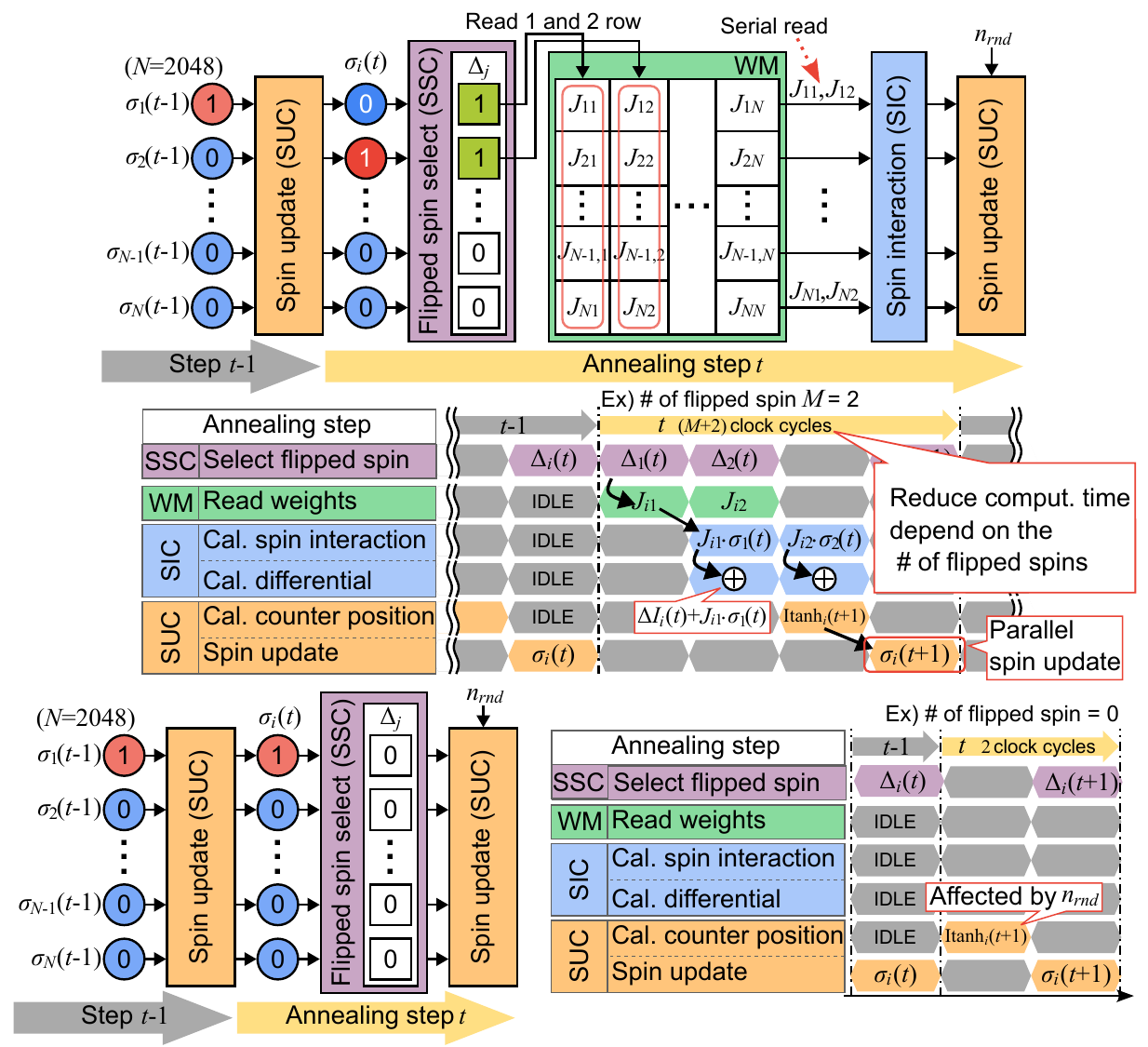}
		\caption{Datapath timing of the DSSA serializer. The trace shows how WM fetches, SIC accumulation, and SUC comparisons overlap when spins flip densely at the start of a shot and then collapse to a few cycles once the SSC queue empties, highlighting the latency advantage of differential scheduling over the fixed-length serialized SSA baseline. The highlighted boxes are shown with solid borders to improve visibility in grayscale printing.}
	\label{fig:datapath}
\end{figure}

\subsection{Spin-Gate and Update Circuits}
Fig.~\ref{fig:spingate} magnifies the SG module and clarifies the per-shot timeline. Each SG comprises a spin interaction circuit (SIC) that accumulates the net differential current $\Delta I_i$ from freshly flipped neighbors according to \eqref{eq:delta_current} and a spin update circuit (SUC) that compares the accumulated value with a stochastic threshold to derive the next state $\sigma_i(t+1)$.

The SUC raises a differential flag $\Delta_i$ whenever $\sigma_i$ flips, enabling the spin-select circuit (SSC) to operate on a sparse vector rather than a dense list of 2,048 entries. To preserve accuracy with modest hardware, the SUC implements the hyperbolic tangent in \eqref{eq:probability} via a stochastic finite-state machine, and the SIC stores the 4-bit interconnection weights broadcast by the WM banks. The timeline beneath Fig.~\ref{fig:spingate} corresponds to the textual description: initialization fills the SIC registers, each flip triggers a DSSA update, and the SSC drains the queue before the scheduler launches the next shot.
The SUC FSM follows the integral stochastic $\tanh(\cdot)$ style: a small up/down counter perturbed by shared random bits realizes the probability in \eqref{eq:probability} without LUTs, and its output bit drives the $\sigma_i$ latch and $\Delta_i$ flag. The SIC taps the WM banks once per active spin, accumulates signed 4-bit weights plus $h_i$, and feeds the FSM input. The inset and timeline in Fig.~\ref{fig:spingate} highlight these signals—shared RNG input, SIC accumulation, SUC comparison, $\Delta_i$ assertion, and SSC queue drain.

\subsection{Stochastic Nonlinearity and RNG Sharing}
DSSA reuses the hyperparameter controls of SSA---$I_{0\min}$, $I_{0\max}$, $n_{\text{rnd}}$, $\tau$, and $\beta$---but interprets them over sparsified activity. In hardware, $n_{\text{rnd}}$ shared random bits per 16-spin group feed the SUC’s stochastic FSM that approximates $\tanh(\cdot)$ with negligible area, matching the 16:1 sharing shown in Fig.~\ref{fig:architecture}. The scheduler cools $I_0$ step-by-step using \eqref{eq:temperature}; one shot is defined as a full ramp from $I_{0\min}$ to $I_{0\max}$, so temperature updates occur every step while the shot counter advances only after a complete ramp.

\subsection{Complexity and Workload Reduction}
During each shot, only spins with $\Delta_i{=}1$ enqueue into the spin-select circuit; the queue drains until no flips remain, at which point the scheduler advances the temperature and launches the next shot. In the initialization shot, every spin’s interaction current is computed to establish a consistent starting point.

Thereafter, the SIC accumulators store weights from a single WM fetch per active spin, and the SUC performs a one-cycle stochastic comparison. By decoupling WM reads from SUC updates and allowing WM to idle when no flips occur, DSSA minimizes wasted toggling. The serialized datapath trades a few cycles per flip for dramatically reduced total cycles versus naive serialization that would revisit all $N$ spins every step, shrinking computation to the active frontier while maintaining correctness.

\subsection{Serializer Timing and Latency}
Latency advantages stem directly from the datapath in Fig.~\ref{fig:datapath}. Only differential contributions from flipped spins are recomputed; interactions from unflipped spins are skipped entirely. DSSA therefore scales latency with the number of flips instead of the total spin count, reducing serialized SSA slow-downs by up to three orders of magnitude for 2,000-spin problems. Early shots, when the solution is highly dynamic, spend more time performing updates, but late shots benefit from sparse activity as the system converges. This data-dependent workload is explicit in the lower traces of Fig.~\ref{fig:datapath}, where the selective update window closes as soon as the SSC runs out of pending flips.
Early traces in Fig.~\ref{fig:datapath} show WM fetch, SIC accumulation, and SUC comparison overlapping for hundreds of cycles when flips are dense, while late shots collapse to a few dozen cycles once the SSC queue is nearly empty, visually tying the latency gain to flip sparsity.
The upper traces also highlight that WM and SIC buses go idle between sparse flips, and the solid fixed-length serialized SSA baseline{} remains flat at the full $N$-cycle cost, underscoring how DSSA’s queue-based scheduler trims both active cycles and wasted toggling when activity is low.

Thus, the $M+2$-cycle behavior in Fig.~\ref{fig:datapath} results from differential update scheduling based on the number of flipped spins $M$, not from a conventional fixed-length time-division schedule over all $N$ spins.

\begin{table}[htbp]
	\centering
	\scriptsize
	\setlength{\tabcolsep}{3pt}
	\small
	\setlength{\tabcolsep}{3.5pt}
	\small
	\setlength{\tabcolsep}{3.5pt}
	\caption{DSSA MAX-CUT results across 100 trials (min/avg/max).
		All results are obtained under fixed annealing conditions:
			5 shots per trial, annealing steps per shot as listed in Table~\ref{tab:hyperparams},{} and hyperparameters
			selected via grid search as described in Section~\ref{subsec:benchmark_setup}.
		$|V|$ and $|E|$ denote the number of vertices and edges of each G-set instance.}
	\label{tab:results_min_mean_max}
	\begin{tabular}{lcccccccc}
		\hline
		Problem &
		$|V|$ &
		$|E|$ &
		Min &
		Min [\%] &
		Avg. &
		Avg. [\%] &
		Max &
		Max [\%] \\
		\hline
		G22 & 2000 & 19990 & 13280 & 99.4 & 13318.0 & 99.7 & 13349 &99.9\\
		G23 & 2000 & 19990 & 13280 & 99.5 & 13309.5 & 99.7 & 13337 &99.9\\
		G24 & 2000 & 19990 & 13283 & 99.6 & 13300.3 & 99.7 & 13318 &99.9\\
		G27 (avg HP) & 2000 & 19990 & 3289 & 98.4 & 3317.1 & 99.3 & 3338 &99.9\\
		G27 (max HP) & 2000 & 19990 & 3285 & 98.3 & 3316.8 & 99.3 & 3341 &100\\
		G35 & 2000 & 11778 & 7606 & 98.9 & 7634.1 & 99.3 & 7651 &99.5\\
		G39 & 2000 & 11778 & 2345 & 97.4 & 2380.8 & 98.9 & 2397 &99.5\\
		K2000 & 2000 & 1999000 & 33073 & 99.2 & 33229.1 & 99.7 & 33337 & 100\\
		\hline
	\end{tabular}
\end{table}

\begin{figure}[htbp]
	\centering
	\includegraphics[width=0.33\textwidth]{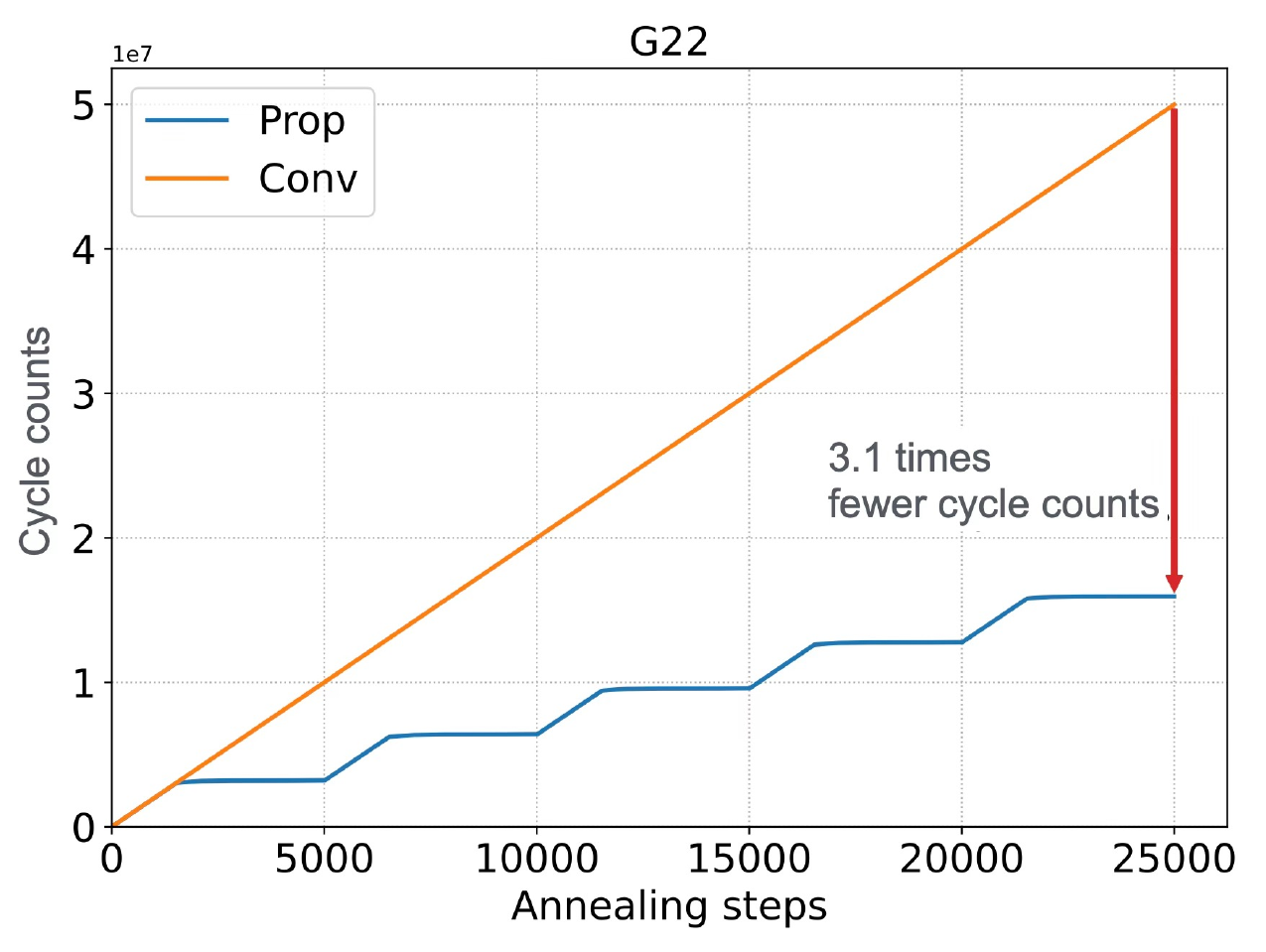}\hfill
	\includegraphics[width=0.33\textwidth]{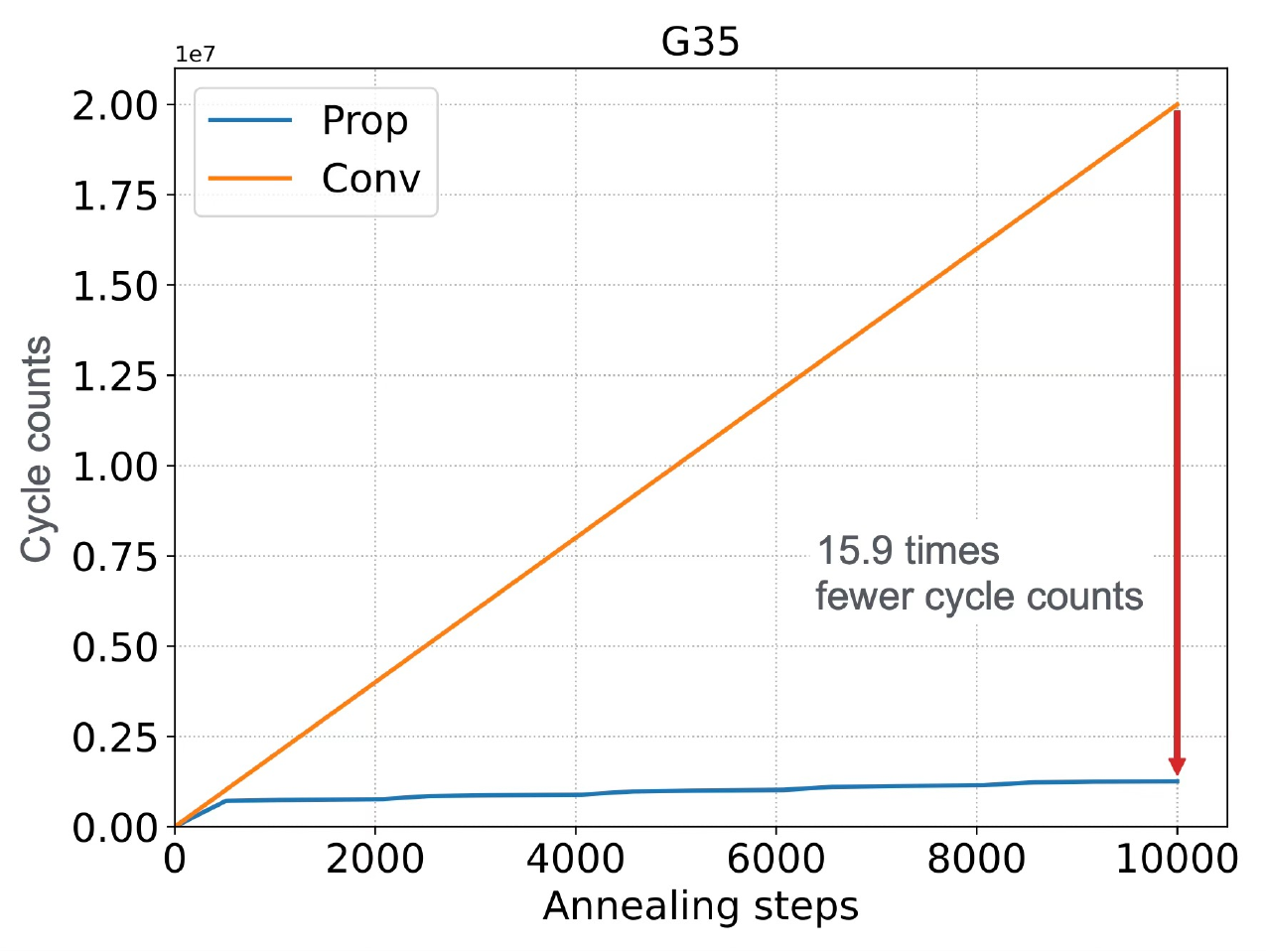}\hfill
	\includegraphics[width=0.33\textwidth]{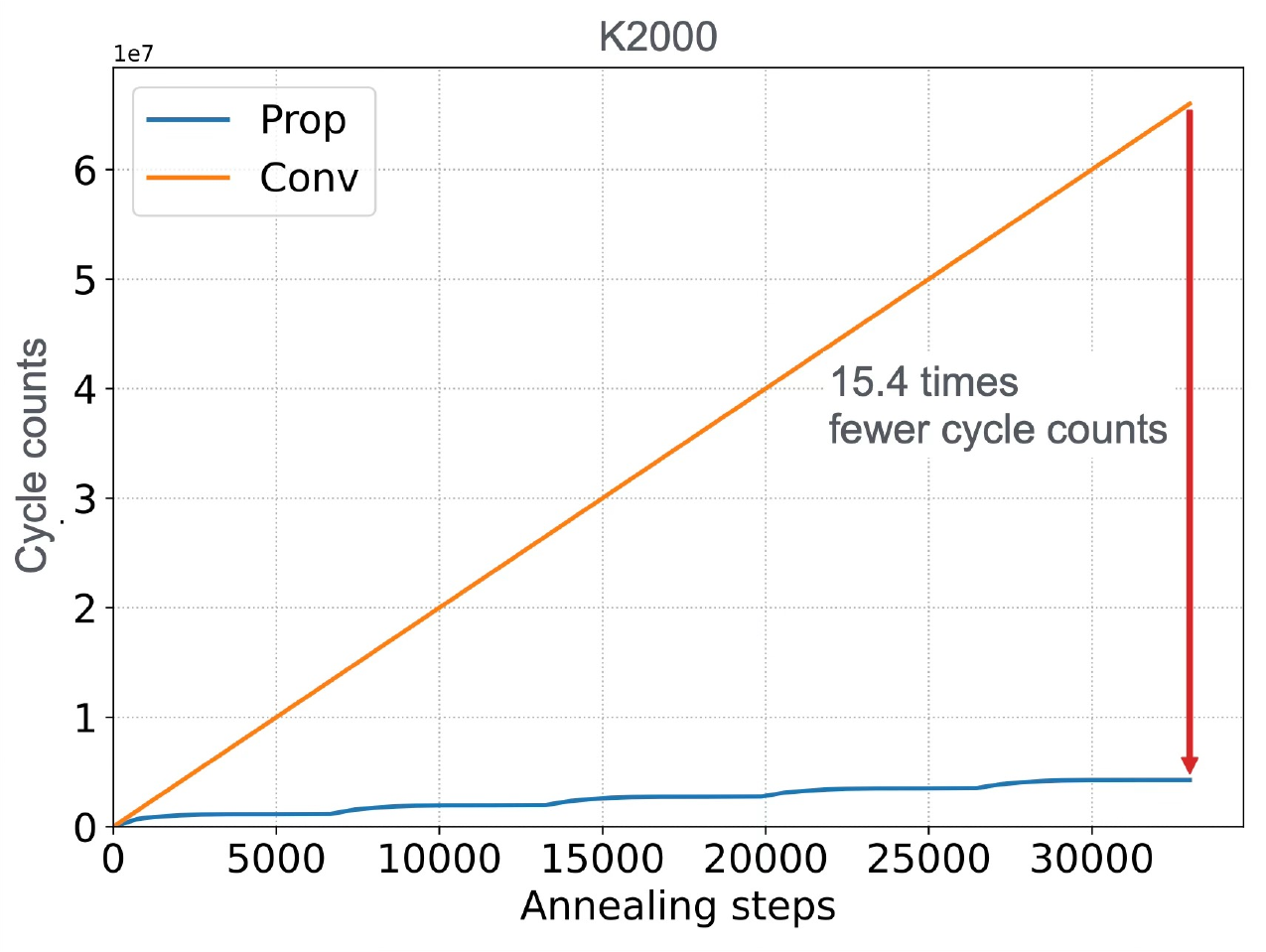}
	\caption{Cumulative cycles/step: proposed DSSA (prop.) vs. serialized SSA (conv.) on G22, G35, and K2000, assuming a fully serialized 2,000-connection baseline. One annealing step evaluates all $N$ spins once; y-axis shows cumulative cycles.
	The stepwise increase observed for the proposed DSSA reflects its
	event-driven execution: cycle counts increase only when spin flips occur.
	When no spins flip, the scheduler advances the annealing step with minimal
	overhead, resulting in flat segments in the cumulative cycle count.
}
	\label{fig:cnt_multi}
\end{figure}

\section{Experimental Evaluation}
\label{sec:evaluation}
\subsection{Benchmark Setup and Hyperparameter Search}
\label{subsec:benchmark_setup}
Software evaluation follows the pre-layout evaluation workflow.
 Simulations run on an Intel Xeon Gold~5318Y (\SI{2.1}{\giga\hertz}, Rocky~Linux~8.6) with a Python~3.8 DSSA implementation.

Seven 2,000-node MAX-CUT instances are exercised: G22, G23, G24, G27 (random graphs, 19{,}990 edges), G35 and G39 (planar graphs, 11{,}778 edges), and K2000 (fully connected, 1{,}999{,}000 edges). G22/23/24/35 use $+1$ edge weights, while G27, G39, and K2000 use $\{\pm1\}$. The best-known cut values are 13{,}359, 13{,}344, 13{,}337, 3{,}341, 7{,}687, 2{,}408, and 33{,}337, respectively. 

The G-set benchmark suite is widely used for evaluating MAX-CUT algorithms and optimization hardware. The selected instances cover different graph characteristics, including random graphs, planar graphs, and the dense fully connected K2000 instance, allowing the proposed DSSA architecture to be evaluated under different connectivity and weight patterns.

To further evaluate the general applicability of the proposed architecture beyond the G-set benchmark suite, we additionally evaluated seven MAX-CUT instances from the Biq Mac Library~\cite{Wiegele2007}. 
These benchmarks were selected from the Biq Mac Library, an independent benchmark collection containing graph instances with diverse structural characteristics and edge-weight distributions that is widely used for evaluating MAX-CUT and quadratic binary optimization algorithms.
Since prior ASIC annealing processors rarely report results on Biq Mac instances, these experiments are intended to validate the generality of the proposed DSSA solver rather than to provide direct hardware-to-hardware comparisons.


Prior to full-length anneals, a grid search tunes DSSA hyperparameters: $I_{0\min}\!\in\![1,4]$, $I_{0\max}\!\in\![2^5,2^{10}]$, $n_{\text{rnd}}\!\in\![1,32]$, and $\tau\!\in\![100,600]$, with $\beta{=}1$ fixed to shorten exploration.
%
For each hyperparameter tuple, five trials of five shots are executed;
the average and best scores select per-benchmark settings. Because DSSA skips spins that do not flip after initialization, computation stays proportional to the sparse flip activity typically seen in later shots despite the 2,000-spin scale.

The benchmark mix spans dense (K2000), random (G22/23/24/27), and planar (G35/39) graphs to cover connectivity and weight patterns seen in prior SSA studies. Hyperparameters are fixed per benchmark after the grid search above; they are not retuned across runs, keeping comparisons consistent for the cycle-count and TTS analyses that follow.
The full min/average/max cut statistics for these instances are summarized in Table~\ref{tab:results_min_mean_max}, and the cumulative cycle behavior used to quantify runtime reduction is shown in Fig.~\ref{fig:cnt_multi}.

Although DSSA can target other COPs up to 2,048 nodes (for example, TSP or graph coloring), this paper focuses on MAX-CUT to align with benchmarks used in prior annealing hardware studies and to enable direct comparison with recent fully connected chips.

\subsection{Design and Implementation Flow}
\label{subsec:design_flow}
The DSSA processor RTL is written in SystemVerilog and synthesized with Synopsys Design Compiler U-2022.12. Because the design is fully digital, TSMC \SI{28}{\nano\meter} standard-cell and I/O libraries are used throughout logic synthesis. The synthesized gate-level netlist is placed and routed in Cadence Innovus v21.12, which handles SRAM macro placement, I/O cell placement, and automatic P\&R of the gate-level design. All hardware results reported in this section are derived from post-layout timing and power evaluation unless stated otherwise.

In addition to the ASIC post-layout evaluation, we verified the RTL implementation through FPGA place-and-route on an AMD/Xilinx Virtex UltraScale+ VCU118 evaluation platform  using Vivado 2022.2 with a \SI{150}{\mega\hertz} timing constraint. The design successfully completed implementation, confirming the synthesizability and hardware feasibility of the proposed DSSA architecture. The FPGA resource utilization was 198{,}586 LUTs (16.8\%), 78{,}559 FFs (3.32\%), 480 BRAMs (22.2\%), 102 I/Os (12.3\%), and 3 BUFGs (0.17\%).

The FPGA place-and-route result is used only to confirm RTL synthesizability and implementation feasibility. It is not used as a substitute for ASIC silicon measurement, and all reported ASIC timing, area, and power results are based on the post-layout design flow described above.

	\begin{table}[htbp]
		\centering
		\scriptsize
		\setlength{\tabcolsep}{2.5pt}
		\caption{Hyperparameters used for DSSA evaluation.
			For G27, two parameter sets are used: ``avg HP'' corresponds to the
			hyperparameters selected by average-score criterion, while ``max HP''
			corresponds to those selected by best-score criterion.
			Steps/shot is computed as $N_I\tau$, where $N_I=\lfloor\log(I_{0,\max}/I_{0,\min})/\log(2^\beta)\rfloor+1$.}
	\label{tab:hyperparams}
		\begin{tabular}{lcccccc}
			\hline
			Problem & $n_{\text{rnd}}$ & $I_{0,\min}$ & $I_{0,\max}$ & $\beta$ & $\tau$ & Steps/shot \\
			\hline
			G22 & 7 & 1 & 512 & 1 & 500 & 5{,}000 \\
			G23 & 4 & 4 & 64 & 1 & 600 & 3{,}000 \\
			G24 & 7 & 1 & 256 & 1 & 600 & 5{,}400 \\
			G27 (avg HP) & 3 & 1 & 32 & 1 & 300 & 1{,}800 \\
			G27 (max HP) & 3 & 1 & 2048 & 1 & 300 & 3{,}600 \\
			G35 & 3 & 4 & 32 & 1 & 500 & 2{,}000 \\
			G39 & 3 & 2 & 256 & 1 & 600 & 4{,}800 \\
			K2000 & 24 & 1 & 1024 & 1 & 600 & 6{,}600 \\
			\hline
		\end{tabular}
	\end{table}

\begin{table}[htbp]
	\caption{Additional DSSA MAX-CUT results on seven Biq Mac Library instances across 100 trials. The gap is computed relative to the best-known cut value. All results were obtained using the common hyperparameter set C defined in Section~\ref{subsec:design_flow}.}
	\label{tab:biqmac_results}
	\centering
	\scriptsize
	\setlength{\tabcolsep}{3.5pt}
	\begin{tabular}{lrrrrrc}
		\hline
		Problem & Best-known & Avg. & Max & Avg. gap & Avg. gap [\%] & Best-known reached \\
		\hline
		g05\_100.0 & 1430 & 1421.03 & 1430 & 8.97 & 0.627 & Yes \\
		pm1d\_100.0 & 340 & 340.00 & 340 & 0.00 & 0.000 & Yes \\
		pm1s\_100.0 & 127 & 126.62 & 127 & 0.38 & 0.299 & Yes \\
		pw05\_100.0 & 8190 & 8119.26 & 8190 & 70.74 & 0.864 & Yes \\
		w01\_100.0 & 651 & 644.40 & 651 & 6.60 & 1.014 & Yes \\
		w05\_100.0 & 1646 & 1609.83 & 1635 & 36.17 & 2.197 & No \\
		w09\_100.0 & 2121 & 2037.94 & 2121 & 83.06 & 3.916 & Yes \\
		\hline
	\end{tabular}
\end{table}

To examine the sensitivity to benchmark-specific tuning, we further evaluated four common hyperparameter sets across all benchmarks and compared them with the benchmark-specific parameters in Table~\ref{tab:hyperparams}. The common candidates are A: $n_{\mathrm{rnd}}=3$, $I_{0\min}=1$, $I_{0\max}=256$, $\beta=1$, $\tau=500$; B: $n_{\mathrm{rnd}}=4$, $I_{0\min}=1$, $I_{0\max}=256$, $\beta=1$, $\tau=600$; C: $n_{\mathrm{rnd}}=7$, $I_{0\min}=1$, $I_{0\max}=512$, $\beta=1$, $\tau=600$; and D: $n_{\mathrm{rnd}}=3$, $I_{0\min}=2$, $I_{0\max}=256$, $\beta=1$, $\tau=500$.

\begin{figure}[htbp]
	\centering
	\includegraphics[width=0.82\textwidth]{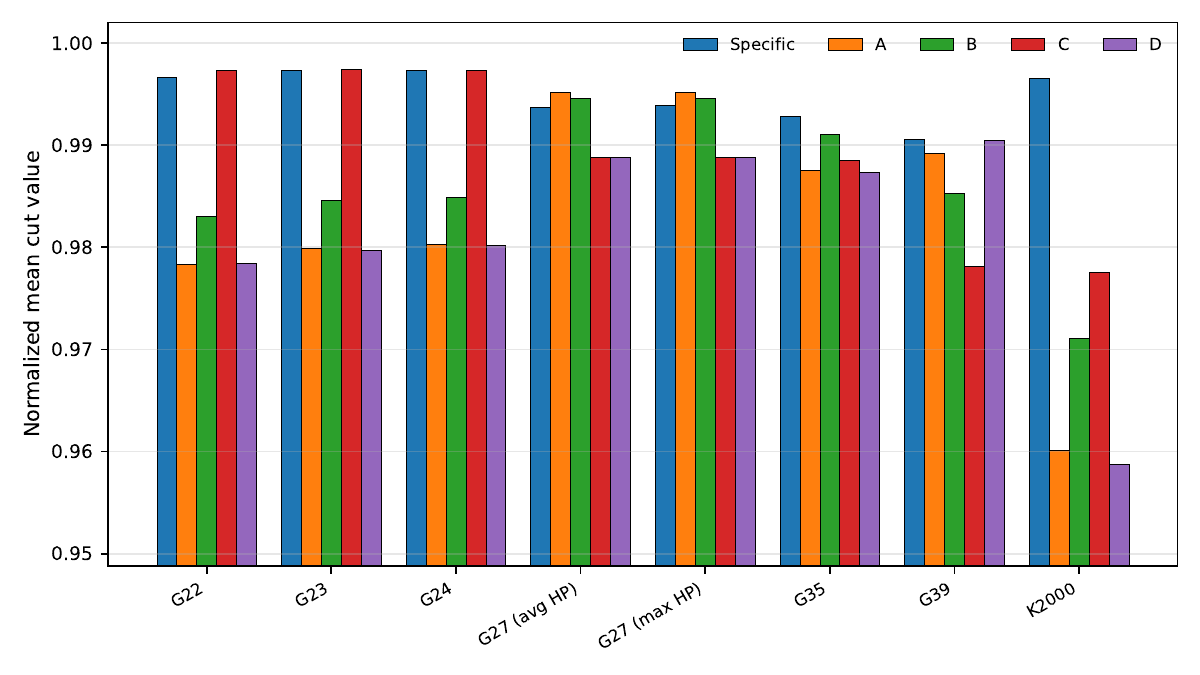}
	\caption{Hyperparameter dependence of DSSA across benchmark-specific and common parameter settings. The vertical axis shows the normalized mean cut value relative to the best-known cut value for each benchmark. ``Specific'' denotes the benchmark-specific parameters in Table~\ref{tab:hyperparams}, while A–D correspond to the four common parameter sets defined in Section~\ref{subsec:design_flow}.}
	\label{fig:hyperparameter_dependence}
\end{figure}

The comparison in Fig.~\ref{fig:hyperparameter_dependence} indicates that common parameter settings can provide useful solution quality for multiple graph structures, while benchmark-specific tuning or selecting a suitable common candidate can further improve robustness. This observation also suggests that lightweight online adaptation based on runtime indicators, such as flip activity, could be considered in future work.

\begin{table}[htbp]
	\caption{Serialized SSA vs. DSSA cycle counts (2,000-spin fully connected assumption).}
	\label{tab:cnt_comp}
	\centering
	\resizebox{\textwidth}{!}{%
	\begin{tabular}{l|c|c|c|c}
		\hline\hline
		Problem & Cycles/step SSA & Cycles/step DSSA & Total speedup & Avg. flips/step (DSSA, /2{,}000) \\
		\hline
		G22 & 2{,}000 & 638.0 & 3.1$\times$ & 31.9\% \\
		G23 & 2{,}000 & 186.2 & 10.7$\times$ & 9.3\% \\
		G24 & 2{,}000 & 707.3 & 2.8$\times$ & 35.4\% \\
		G27 (avg HP) & 2{,}000 & 94.6 & 21.1$\times$ & 4.7\% \\
		G27 (max HP) & 2{,}000 & 47.5 & 42.1$\times$ & 2.4\% \\
		G35 & 2{,}000 & 126.0 & 15.9$\times$ & 6.3\% \\
		G39 & 2{,}000 & 65.4 & 30.6$\times$ & 3.3\% \\
		K2000 & 2{,}000 & 129.6 & 15.4$\times$ & 6.5\% \\
		\hline\hline
	\end{tabular}%
	}
\end{table}

\subsection{Solution Quality and Runtime Reduction}

Table~\ref{tab:results_min_mean_max} summarizes DSSA MAX-CUT results across 100 trials (min/avg/max).
Each trial consists of five shots under fixed annealing conditions.
Average cuts exceed 99\% of the best-known values for every benchmark; G22/G23/G24/K2000 also clear 99\% on the worst trial, and K2000 plus G27 reach the optimum.
The hyperparameters used for each benchmark are listed in
Table~\ref{tab:hyperparams}.
{}

Table~\ref{tab:biqmac_results} reports the additional evaluation results on seven Biq Mac Library instances. 
All Biq Mac experiments were performed using the common hyperparameter set C, which showed consistently strong performance across the G22, G23, G24, and K2000 benchmarks.
DSSA reached the best-known cut value in six out of seven instances within 100 trials. For the remaining instance, w05\_100.0, the best DSSA result was 1635, which is only 0.668\% below the best-known value of 1646. These results indicate that DSSA maintains high solution quality beyond the G-set benchmark suite.
The Biq Mac experiments did not directly measure cycle reduction; instead, their high solution quality supports the applicability of the DSSA solver beyond the G-set benchmark family. The cycle reduction is governed by flip sparsity, or equivalently the active-frontier size $M$, rather than by the graph family itself. Therefore, the maximum $30.6\times$ total-cycle reduction observed in Table~\ref{tab:cnt_comp} is benchmark- and activity-dependent and is not guaranteed for every problem instance.

DSSA’s cycle count shrinks because only flipped spins recompute interactions. Fig.~\ref{fig:cnt_multi} aggregates representative cumulative cycles/step assuming a fully serialized 2,000-connection baseline for G22, G35, and K2000; the DSSA curves track flip activity and flatten quickly, while the conventional serialized SSA remains linear in $N$. Table~\ref{tab:cnt_comp} reports per-step cycle averages and total-cycle speedups: DSSA cuts cycles by $2.8\times$--$42\times$ per step and $3.1\times$--$30.6\times$ overall. For K2000, only 6.5\% of spins flip per step on average, yielding a $15.4\times$ total-cycle reduction. The fully serialized SSA baseline is intentionally simple to highlight DSSA’s benefit; more aggressive serializers (for example, dual-ported WM or limited parallelism) 
would change absolute cycle counts but would still require $O(N)$ work
each step, whereas DSSA scales with the much smaller $M$ flipped spins. Unlike conventional serialized SSA, DSSA does not incur a fixed number of cycles at every annealing step. Instead, cycle consumption is proportional to the number of flipped spins. As a result, periods with sparse flip activity appear as flat regions in Fig.~\ref{fig:cnt_multi}, while steps with clustered spin flips produce abrupt increases, yielding the observed stepwise behavior.{}
\begin{figure}[htbp]
	\centering
	\includegraphics[width=\columnwidth]{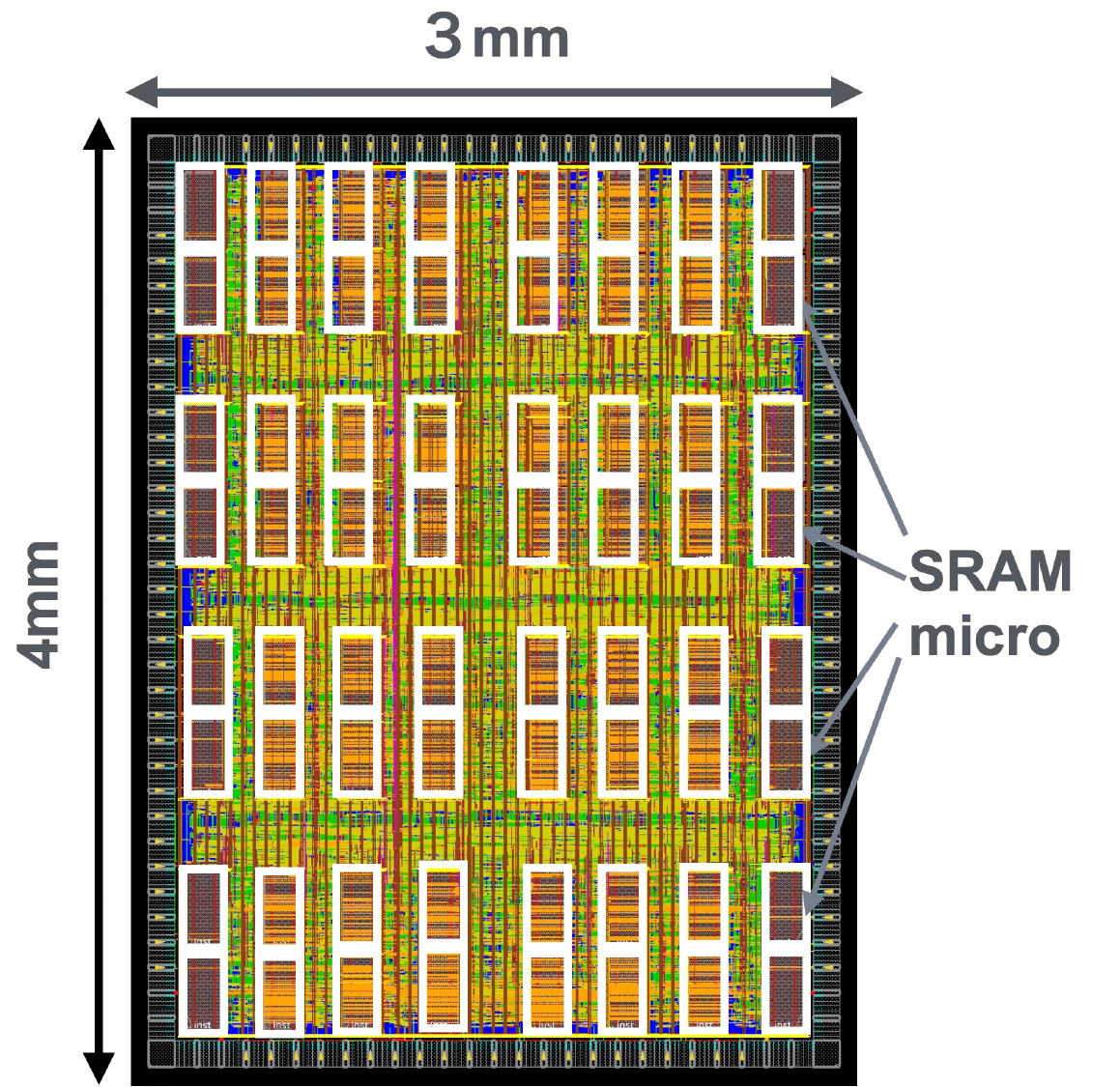}
	\caption{Layout of the 2,048-spin fully connected DSSA processor in TSMC \SI{28}{\nano\meter}.}
	\label{fig:layout}
\end{figure}

The post-layout floorplan in Fig.~\ref{fig:layout} provides the physical view corresponding to the distributed SRAM-macro organization discussed in Section~\ref{sec:architecture}.

\begin{table}[htbp]
	\caption{Overview comparison of hardware characteristics of recent annealing systems and the proposed DSSA processor.
		This table focuses on hardware-level metrics such as process, scalability, and power, while detailed performance comparisons on benchmark instances are provided in Table~\ref{tab:k2000_tts}.
		For the ISSCC'21 King's-graph entry, power and power/spin are marked as N/A because they are not reported in the cited work.}
	\label{tab:overview_comp}
	\centering
	\scriptsize
	\setlength{\tabcolsep}{3pt}
	\resizebox{\textwidth}{!}{%
	\begin{tabular}{lccccccc}
		\hline\hline
		Work & Process & Spins & Topology & Clock & Power & Power/spin & Validation \\
		\hline
		ISSCC'21~\cite{Takemoto2021ISSCC} & 40\,nm & 16k ($\times$9 chips) & King's graph (sparse) & 100 MHz & N/A & N/A & Silicon \\
		JSSC'21~\cite{Yamamoto2021JSSC} & 65\,nm & 512 & Fully connected & 320 MHz & 649 mW & 1.27 mW & Silicon \\
		STATICA~\cite{Yamamoto2021JSSC} (proj.) & 65\,nm & 2000 & Fully connected & 320 MHz & $2.0\times10^3$ mW & 1.00 mW & Projected silicon \\
		ISSCC'23~\cite{Kawamura2023ISSCC} & 40\,nm & 512 ($\times$4 chips) & Fully connected & 134--336 MHz & 151.6--474.9 mW & 0.29--0.93 mW & Silicon \\
		This work & 28\,nm & 2048 & Fully connected & 500 MHz & 316.1 mW & 0.15 mW & Post-layout simulation \\
		\hline\hline
	\end{tabular}%
	}
\end{table}

\begin{table}[htbp]
	\caption{Detailed performance and energy comparison on the K2000 MAX-CUT instance.
		For the GPU baseline, reported runtime and power include all computation and
		memory access overheads. For the proposed DSSA processor, the reported values
			cover only the annealing phase; setup steps such as spin initialization and
			temperature scheduling are excluded. The SSA baseline follows the stochastic simulated
			annealing algorithm described in~\cite{SSA}.
				The energy per spin per step is computed using $N=2000$ and $N_{\mathrm{step}}=9000$; this metric is not reported for projected STATICA because the same step-level definition is not directly available.
			}
	\label{tab:k2000_tts}
	\centering
	\begin{tabular}{l|ccc}
		\hline\hline
		& GPU SSA & STATICA~\cite{Yamamoto2021JSSC}(proj.) & This work \\
		\hline
		Power [mW] & $52.8\times 10^3$ & $2.0\times 10^3$ & 316.1 \\
		Runtime $T$ [ms] & 499.88 & 0.48 & 2.30 \\
		$P_a(T)$ & 0.45 & 0.77 & 0.98 \\
		$\mathrm{TTS}(0.99)$ [ms] & 3{,}850.6 & 1.5 & 2.7 \\
		Energy at & $203.3\times 10^3$ & 3.0 & 0.86 \\
		$\mathrm{TTS}(0.99)$ [mJ]\\
			Energy/spin/step & \SI{1.47}{\micro\joule} & -- & \SI{40.4}{\pico\joule} \\
		\hline\hline
	\end{tabular}
\end{table}

\subsection{Ablation: RNG Sharing and Differential Serialization}

\begin{figure}[htbp]
	\centering
	\includegraphics[width=\columnwidth]{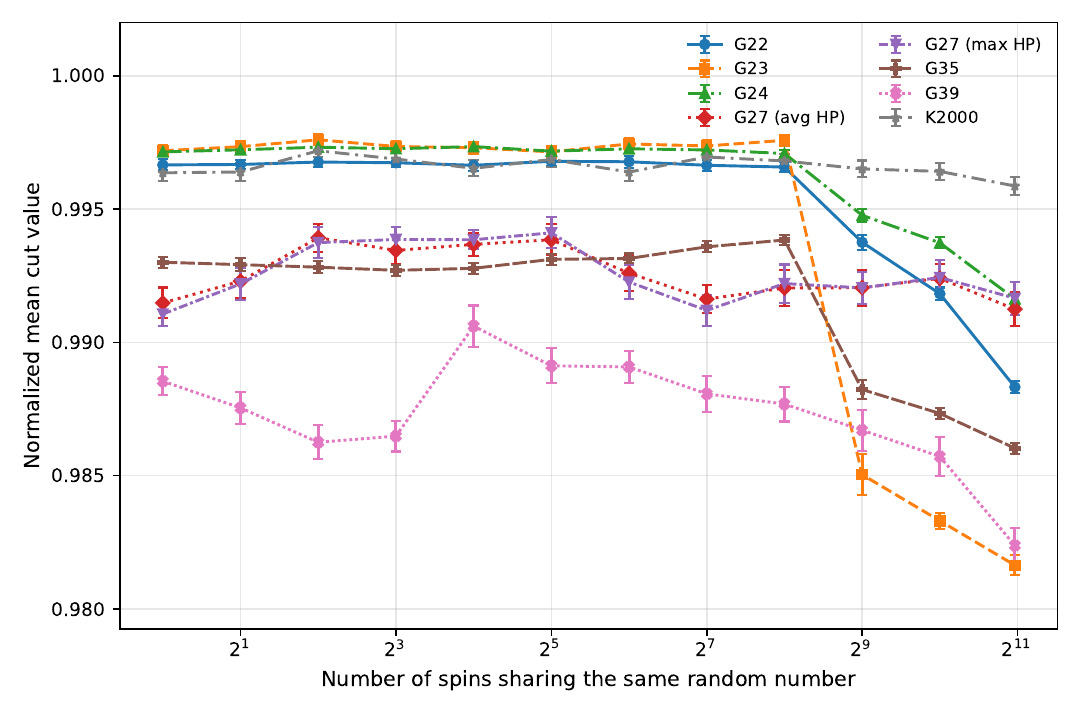}
	\caption{Impact of RNG sharing on normalized mean cut value across multiple random seeds. Error bars indicate 95\% confidence intervals over 100 trials with random seeds 0--99. The results show that 16:1 RNG sharing maintains solution quality within a narrow statistical range while reducing RNG hardware cost.}
	\label{fig:rng_sharing_stats}
\end{figure}

Fig.~\ref{fig:rng_sharing_stats} summarizes additional simulations for the seven G-set benchmarks used in this paper under different RNG sharing factors and across multiple random seeds, reporting the normalized mean cut value relative to the best-known cut value. For each benchmark and RNG-sharing factor, 100 trials with random seeds 0--99 were performed, with five shots per trial. Error bars indicate 95\% confidence intervals. The narrow confidence intervals indicate that the effect of RNG sharing is stable across seeds under the evaluated conditions; this work adopts 16 spins per shared RNG, resulting in an eightfold improvement in RNG area efficiency compared to a non-shared design.

Excessive sharing reduces entropy diversity across spins, degrading exploration.
{}{}

Disabling the differential scheduler (naive serialization) restores the $N^2$ interaction cost and erases the $3.1\times$--$30.6\times$ cycle reductions in Table~\ref{tab:cnt_comp}. Both choices are essential to reach the reported density and speed targets.

\subsection{Hardware Validation and Comparison}

The proposed processor targets TSMC \SI{28}{\nano\meter} and integrates 2,048 fully connected spins, a 16\,Mb 4-bit WM built from sixty-four 2{,}048$\times$128 SRAM macros, and clock gating for sparse updates. The die measures \SI{3}{\milli\meter} $\times$ \SI{4}{\milli\meter}; post-layout utilization is 35.7\% with a gate count of 8{,}498{,}667.

The design closes timing at \SI{500}{\mega\hertz} (synthesis limited at \SI{600}{\mega\hertz}) with \SI{0.9}{\volt} core supply and \SI{1.8}{\volt} I/O. 
The estimation assumes TT at \SI{25}{\celsius} with $V_{\mathrm{core}}=0.9$\,V and $V_{\mathrm{IO}}=1.8$\,V.{}{} 
Estimated power is \SI{316.1}{\milli\watt} at \SI{500}{\mega\hertz}, or \SI{0.15}{\milli\watt} per spin. The power breakdown is \SI{223.85}{\milli\watt} (clock), \SI{20.30}{\milli\watt} (memory), and \SI{71.95}{\milli\watt} (others).{}{} The package uses 118 pins, including dedicated rails for WM and logic domains; the resulting floorplan is shown in Fig.~\ref{fig:layout}.

In the post-layout floorplan, the 64 SRAM macros are distributed around the spin-gate array to shorten the weight-delivery paths and to balance routing congestion. The spin-select circuit, priority encoder, serializer, and multiplexer network are placed near the interface between the spin-gate array and the weight-memory banks so that selected flipped-spin indices can drive sequential WM accesses with reduced long-distance routing. This physical organization realizes the differential-update datapath shown in Fig.~\ref{fig:datapath}: only the weight rows associated with selected flipped spins are read and accumulated, while unchanged-spin contributions are retained locally.

The FPGA result is used only to verify RTL synthesizability and implementation feasibility, whereas the reported performance and energy results are based on ASIC post-layout evaluation.

Table~\ref{tab:overview_comp} provides an overall view of recent annealing hardware~\cite{Takemoto2021ISSCC,Yamamoto2021JSSC,Kawamura2023ISSCC}. The comparison highlights that prior fully connected chips either cap spins at 512 or require multi-chip tiling, whereas DSSA keeps 2,048 fully connected spins and all weights on a single die. Despite the on-chip memory, the \SI{12}{\milli\meter\squared} area is comparable to smaller-spin designs. The \SI{500}{\mega\hertz} clock is the highest among the fully connected entries, and the per-spin power of \SI{0.15}{\milli\watt} is 8.5$\times$ lower than the JSSC'21 fully connected chip at similar $V_{\text{DD}}$ and below the 2023 metamorphic design even when normalized to frequency.

Table~\ref{tab:overview_comp} focuses on hardware-level characteristics, whereas detailed performance comparisons on benchmark instances are provided separately in Table~\ref{tab:k2000_tts}. Validation distinguishes fabricated silicon, projected silicon, and post-layout simulation, which should be considered when interpreting cross-work comparisons.

Note that the interaction coefficient precision differs across designs. The ISSCC'21 annealer employs fixed binary (1-bit) interactions on a sparse King's graph topology, whereas STATICA and the proposed DSSA both use 4-bit multilevel interaction coefficients for fully connected graphs. Amorphica supports programmable interaction coefficients, although the exact bit width is not explicitly specified in the published description. Therefore, the power comparison in Table~\ref{tab:overview_comp} should be interpreted in light of these differences. Importantly, the proposed DSSA achieves lower power per spin despite using comparable interaction precision to prior fully connected designs, indicating that the reported power reduction primarily originates from architectural efficiency rather than reduced coefficient precision.
{}

Time-to-solution (TTS) quantifies annealing efficiency for the 2,000-node K2000 instance. For target success probability $P$,
\begin{equation}
 \mathrm{TTS}(P) = T \cdot \frac{\ln(1-P)}{\ln\!\left(1-P_a(T)\right)},
 \label{eq:tts}
\end{equation}
where $T$ is the runtime of one anneal and $P_a(T)$ is its success probability. While Table~\ref{tab:overview_comp} gives the cross-work overview, Table~\ref{tab:k2000_tts} retains a focused comparison on the K2000 instance against a GPU SSA implementation (NVIDIA RTX~2080~Ti) and a projected 2,000-spin STATICA extrapolated from the 512-spin chip in~\cite{Yamamoto2021JSSC} under a four-chip tiling assumption. The GPU delivers long runtime and modest $P_a$ but consumes $52.8$\,W; STATICA runs fastest but at lower success probability. DSSA balances both, reaching $P{=}0.99$ with $0.86$\,mJ---$3.5\times$ lower energy than projected STATICA despite a longer runtime, and five orders of magnitude below the GPU baseline. The combination of high $P_a(T)$ and moderate $T$ yields the lowest energy at $\mathrm{TTS}(0.99)$ among the three, illustrating the benefit of sparse differential updates even at a higher clock.

The sparse, parallel-update King's-graph machine and the fully connected DSSA processor therefore occupy different design spaces and target different problem classes. Sparse parallel machines can achieve very short absolute TTS by exploiting restricted connectivity and massive parallel updates, whereas the proposed DSSA targets dense fully connected Ising formulations with multilevel interaction weights stored on chip. Consequently, absolute TTS alone should not be interpreted as a topology-independent ranking across these architectures; the topology, connectivity, coefficient precision, validation methodology, and target problem formulation must be considered together.

The GPU-based SSA baseline executes a conventional SSA algorithm~\cite{SSA}, where runtime and power consumption include all
associated costs such as global memory access and control overhead. For the
proposed DSSA processor, the reported runtime and power cover only the
annealing phase; setup steps such as spin initialization and temperature
scheduling are excluded.
Here, setup refers to the initialization of spin states and loading of
hyperparameters prior to the annealing process.
{}

The GPU baseline was implemented using a CuPy RawKernel-based CUDA kernel. Each CUDA thread updates one spin, and the kernel is launched with 32 threads per block along the spin dimension. For each annealing step, each thread computes the local field by sequentially accumulating all $N$ interaction terms from the dense weight matrix. The implementation does not employ shared-memory tiling; the dense interaction matrix is read directly from global memory. The weight-matrix accesses are not fully memory-coalesced because neighboring threads access different matrix rows, whereas the spin vector is commonly read across threads and can benefit from cache reuse. Runtime was measured using CUDA events around the annealing kernel invocation.

To compare computational energy efficiency under a unified metric, we also compute the energy per spin per step as
\begin{equation}
 E_{\mathrm{spin,step}} = \frac{P T}{N N_{\mathrm{step}}},
 \label{eq:energy_spin_step}
\end{equation}
where $P$ is the power consumption, $T$ is the runtime per anneal, $N$ is the number of spins, and $N_{\mathrm{step}}$ is the number of annealing steps. In this evaluation, $N=2000$ and $N_{\mathrm{step}}=9000$ for both GPU SSA and DSSA.

The added energy-per-spin-per-step metric indicates the energy efficiency of the actual per-step spin-update computation under the same $N$ and step-count assumptions. Because the GPU baseline is a straightforward conventional SSA implementation rather than a highly optimized GPU annealer, this comparison is intended to provide a transparent reference point rather than an upper bound on GPU performance.

\subsection{Scalability and Applicability to Other Problem Classes}
\label{subsec:scalability}

The scalability of the proposed DSSA processor is primarily constrained by the
weight memory required for fully connected Ising models, which scales as
$O(N^2)$. This requirement is fundamental to dense formulations and is shared
by prior fully connected annealers. In contrast, the spin-gate array, scheduler,
and control logic scale linearly with the number of spins $N$.

For example, the present 2,048-spin design uses a 16 Mb 4-bit weight memory. Under the same fully connected storage model, the required weight-memory capacity increases quadratically with $N$, reaching approximately 64 Mb for 4,096 spins and 256 Mb for 8,192 spins.

These values are first-order projections based on the present 2,048-spin implementation and assume the same 4-bit fully connected weight-storage organization. Under this assumption, the projected chip area and leakage power are expected to be dominated by the quadratic growth of the weight memory, whereas the spin-gate array, scheduler, and control logic scale approximately linearly. Therefore, the presented values indicate expected scaling trends rather than implemented results for larger processor sizes.

In terms of runtime behavior, DSSA differs from conventional serialized SSA by
scaling the number of interaction updates with the number of flipped spins $M$,
rather than with $N$. As shown in Fig.~\ref{fig:cnt_multi} and Table~\ref{tab:cnt_comp},
 $M$ remains a small fraction
of $N$ for large MAX-CUT instances, particularly in later annealing stages.
As a result, the effective cycle count per annealing step exhibits sublinear
scaling in practice, even though the underlying problem is fully connected.

Power consumption similarly benefits from sparse activity. Weight memory access
and interaction accumulation are triggered only when spins flip, allowing both
dynamic power and switching activity to track the data-dependent workload.
When flip activity becomes sparse, DSSA avoids unnecessary memory reads and
computation, sustaining large problem sizes without incurring the full
$O(N^2)$ cost at every step.

The proposed architecture is applicable to a broad class of combinatorial
optimization problems that can be mapped to Ising or QUBO formulations,
including MAX-CUT, the Traveling Salesman Problem, and graph coloring.
From an architectural standpoint, no modification is required as long as the
problem can be expressed using pairwise interactions.
However, the effectiveness of DSSA depends on the temporal sparsity of spin
flips during annealing. For problems dominated by strong global constraints,
such as tightly constrained graph coloring, flip activity may remain high for
longer durations. As the flip rate increases, the active-frontier size $M$
grows, requiring more weight-memory accesses and interaction updates and
progressively reducing the cycle advantage of DSSA. In the worst case where
$M \approx N$, DSSA approaches the cycle count of conventional serialized SSA
without loss of correctness.

In addition to the G-set benchmarks, the additional Biq Mac Library results in Table~\ref{tab:biqmac_results} further demonstrate that the proposed DSSA datapath is not specific to a particular MAX-CUT benchmark suite. Although these results are not used for direct comparison with prior ASIC annealers because comparable hardware results are unavailable, they support the applicability of DSSA to different MAX-CUT benchmark collections.

Although the proposed processor targets fully connected Ising models, the quadratic memory growth can be alleviated for sparse or structured optimization problems by exploiting sparse encoding, structured sparse matrices, block-sparse weight storage, weight compression, and hierarchical or external memory organizations. These techniques can reduce the effective memory footprint when many interaction coefficients are zero or when the coefficient matrix exhibits exploitable structure. However, for arbitrary dense fully connected Ising problems, all pairwise interaction coefficients must still be represented explicitly or implicitly, and the $O(N^2)$ memory complexity therefore remains an inherent worst-case limitation.

Exploring adaptive scheduling policies or hybrid update strategies that better
exploit problem-specific structure remains an important direction for future
work.

\subsection{Limitations}
Although post-layout simulation provides a detailed estimate of timing, area, and power after placement and routing, it does not replace fabricated-silicon measurement. In particular, process variation, supply noise, clock jitter, on-chip noise, IR-drop effects, clock-distribution non-idealities, and realistic memory-access behavior may cause measured silicon performance and power to deviate from the reported estimates. Therefore, the reported results should be interpreted as post-layout estimates rather than measured chip results. Fabricating and measuring silicon across process, voltage, and temperature conditions remains an important direction for future work.

\subsection{Discussion}

DSSA preserves the stochastic acceptance rule of conventional SSA while reducing redundant interaction recomputation through differential updates and event-driven scheduling. 
Importantly, DSSA does not modify the probabilistic state-transition rule itself; rather, it reorganizes when and how interaction terms are evaluated.

\textit{Equivalence Conditions.} 
The equivalence between DSSA and serialized SSA holds under the following assumptions: 
1) each spin is evaluated exactly once per annealing step; 
2) the local field $I_i(t+1)$ used in the stochastic update is mathematically identical to that obtained by full recomputation of $\sum_j J_{ij}\sigma_j(t)$; and 
3) the stochastic acceptance probability in \eqref{eq:probability} remains unchanged. 
Under these conditions, DSSA implements the same Markov chain as conventional serialized SSA.

The differential update in \eqref{eq:dssa_prov_I} is algebraically equivalent to full recomputation because it accumulates the exact contribution of flipped spins while preserving previously stored interaction terms. 
Therefore, for any spin $i$, the local field $I_i(t+1)$ in DSSA matches that of serialized SSA at the time of probabilistic evaluation. 
Since the acceptance probability depends only on $I_i(t+1)$ and $I_0(t)$, the transition probability $P(\sigma_i(t+1)|\sigma(t))$ is preserved.

\textit{Effect of Spin-Select Scheduling.} 
The SSC and priority-based weight reads only determine the order in which interaction updates are processed. 
They do not alter which spins are probabilistically evaluated, nor do they change the acceptance rule. 
Each spin is still updated once per annealing step according to \eqref{eq:probability}. 
Thus, scheduling reorders computations but does not modify state-transition probabilities.

\textit{Skip Idle Steps.} 
When no spins flip during an annealing step, all local fields remain unchanged. 
In this case, the resulting transition corresponds to a deterministic self-loop in the Markov chain with probability one. 
DSSA avoids redundant recomputation of interaction terms in this scenario, but no valid state transition is removed or altered. 
Therefore, the underlying Markov chain is preserved.

DSSA does not introduce new theoretical convergence guarantees beyond those of classical simulated annealing. 
Rather, it constitutes an exact computational reformulation under the assumptions stated above. 
If, in a pathological worst case, a large fraction of spins flip at every step ($M \approx N$), DSSA naturally degenerates to conventional serialized SSA behavior without loss of correctness.

From a hardware perspective, DSSA introduces additional control structures, including the spin-select circuit and priority encoder. 
Post-layout analysis shows that these components occupy only a small fraction of the total die area, which is dominated by the on-chip weight memory. 
Although DSSA incurs some control overhead, this cost is outweighed by the substantial reduction in weight memory accesses and interaction computations when flip activity is sparse.

\section{Conclusion}
\label{sec:conclusion}
A 2,048-spin fully connected annealing processor based on DSSA has been designed in a TSMC \SI{28}{\nano\meter} process and evaluated using post-layout simulation. Differential serialization and 16:1 RNG sharing shrink the active workload to spin flips, enabling \SI{0.15}{\milli\watt}/spin estimated power at \SI{500}{\mega\hertz}, \SI{2.7}{\milli\second} TTS, and \SI{0.86}{\milli\joule} energy-to-solution for 2,000-spin MAX-CUT. Comparisons against recent ASIC annealers and a GPU SSA indicate lower per-spin power and the best projected TTS energy at $P{=}0.99$, despite full connectivity and on-chip weights. Future work includes expanding WM capacity and serializer bandwidth for larger spin counts, exploring adaptive temperature schedules tunable on chip, and fabricating silicon to validate the post-layout estimates across benchmarks.

The additional Biq Mac Library evaluation further indicates that the proposed DSSA solver maintains high solution quality beyond the G-set benchmark suite.

\section*{Acknowledgment}
 This research was supported in part by KIOXIA Corporation and by a research grant from the Murata Science and Education Foundation.
This work was supported through the activities of VDEC, d.lab, The University of Tokyo, in collaboration with NIHON SYNOPSYS G.K. and Cadence Design Systems.

\bibliographystyle{IEEEtran}
\bibliography{refs}

\end{document}